\documentclass[useAMS,usenatbib,usegraphicx]{mn2e}
\usepackage[english]{babel}
\usepackage{lscape}

\title[Universal size distribution of void filaments]
{The Size Distribution of Void Filaments in a $\Lambda$CDM Cosmology}
\author[Daeseong Park and Jounghun Lee]
{Daeseong Park\thanks{E-mail: pds2001@astro.snu.ac.kr} and Jounghun Lee\thanks
{E-mail: jounghun@astro.snu.ac.kr}\\
Department of Physics and Astronomy, FPRD, Seoul National University,
Seoul 151-747, Korea \\}

\begin{document}
\label{firstpage} \maketitle

\begin{abstract}
The size distribution of mini-filaments in voids has been derived
from the Millennium Run halo catalogs at redshifts $z=0,0.5,1$ and
$2$. It is assumed that the primordial tidal field originated the
presence of filamentary substructures in voids and that the void
filaments have evolved only little, keeping the initial memory of
the primordial tidal field. Applying the filament-finding algorithm
based on the minimal spanning tree (MST) technique to the Millennium
voids, we identify the mini-filaments running through voids and
measure their sizes at each redshift. Then, we calculate the
comoving number density of void filaments as a function of their
sizes in the logarithmic interval and determine an analytic fitting
function for it. It is found that the size distribution of void
mini-filaments in the logarithmic interval, $dN/d\log S$, has an
almost universal shape, insensitive to the redshift: In the
short-size section it is well approximated as a power-law, $dN/d\log
S \approx S$, while in the long-size section it decreases
exponentially as $dN/d\log S \approx \exp(-S^{\alpha})$. 
We expect that the universal size distribution of void filaments may 
provide a useful cosmological probe without resorting to the rms density 
fluctuations.
\end{abstract}


\section{INTRODUCTION}  \label{intro}

In the classical theory of structure formation, it was generally believed that
the gravity is fully responsible for the formation and evolution of the large
scale structure in the universe. Recently, however, it has been realized that
the overall characteristics of the large scale structures cannot be understood
only in terms of the gravitational influence. It has been pointed out that 
the tidal field provides a driving force in establishing the observed large 
scale filamentary structures in the universe \citep{bon-etal96,lee-evr07,
par-lee07b,hah-etal07,ara-etal07}. The cosmic voids are most vulnerable to the 
tidal influence from the surrounding matter distribution due to their extreme 
low-density \citep{sah-sha96,sha-etal04,sha-etal06,lee-par06,par-lee07b}.
The tidal squeezing and distortion effect tends to deviate the void shapes
from spherical symmetry and could sometimes lead even to the collapse and
disappearance of the voids \citep{sah-etal94,sah-sha96,sha-etal04}.

The presence of the filamentary structures in voids marks the most striking
evidence for the strong tidal influence on voids. The anisotropy
in the spatial distribution of void halos is induced by their alignments
with the principal axes of the tidal tensors \citep{pee01}.
Since the void halos evolve very little and most particles remain primordial
in voids \citep{ein06}, the void mini-filaments are pristine, keeping well
the initial memory of the spatial coherence of the primordial tidal effect,
unlike the large scale filaments which also originated from the large-scale
coherence of the primordial tidal field but have undergone highly nonlinear
processes in the subsequent evolution.  Hence, the void filaments are
the most idealistic probe of the primordial tidal field and its effect on
voids. 

The size distribution of void filaments should be a function of the strength 
of the tidal effect and the host void sizes. The maximum size of a void 
filament cannot exceed the size of its host void.  But, even when a host
void has a large size, it would not have long-size filaments if there is
no tidal influence. Here, we attempt to derive the size distributions of the 
void filaments at various redshifts and to explore their statistical 
properties. Before presenting our results, however, we would like to caution 
the readers for the difficulty in using the void filaments as a probe of the 
primordial tidal field. Unlike the galaxy clusters, there is no unique way to 
define voids and their filaments. Recently, the seminal work of 
\citet{col-etal08} compared thoroughly various void-finding algorithms and 
showed clearly that different void finders different void characteristics, 
under-density profiles, void galaxies etc. Thus, our results are likely to 
be dependent on our specific choice of the void-finder.
The organization of this paper is as follows. In \S \ref{void}, we briefly 
describe the void catalog was obtained from the Millennium Run simulations. 
In \S \ref{filament}, we explain how the void filaments are identified from 
the void catalog. In \S \ref{result}, we derive numerically the size 
distributions of void filaments and determine an analytic fitting formula 
for it. In \S \ref{end}, we discuss the final results and assess the 
future work.

\section{OVERVIEW OF THE VOID IDENTIFICATION}
\label{void}
In our previous work \citep{par-lee07b}, we have obtained a sample
of voids at four different redshifts, $z=0,\ 0.5,\ 1$ and $2$, from
the Millennium Run Simulation \citep{spr-etal05} for a flat
$\Lambda$CDM cosmology with the key cosmological parameters
$\Omega_{m}=0.25$, $\Omega_{b}=0.045$, $h=0.73$,
$\Omega_{\Lambda}=0.75$, $n=1$, and $\sigma_{8}=0.9$, which we use
throughout this paper. Here, we briefly summarize the
void-identification process as an overview.

\begin{figure}
\begin{center}
\includegraphics[width=84mm]{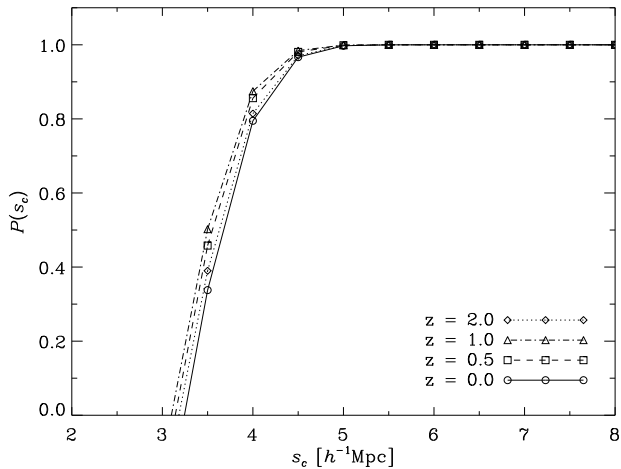}
\caption{The statistical significance $P(s_{c})$ vs. the minimum
void-size threshold $s_{c}$ at $z=0,\ 0.5,\ 1$ and $2$.}
\label{fig:significance}
\end{center}
\end{figure}
We first selected those halos in the Millennium data which consist of more
than $50$ particles since those halos composed of less than $50$ halos 
are statistically unreliable (in private communication with V. Springel). 
Beside, it is interestingly found that when the particle number threshold 
of $N_{c}=50$ is used, the statistical properties of the halo voids are 
quite similar to that of the galaxy voids from the Millennium Sloan Digital 
Sky Survey (SDSS) mock catalog with the magnitude completeness limit.
We applied the void-finding algorithm developed by 
\citet[][hereafter HV02]{hoy-vog02} to the sample of the selected halos
at each redshift, separately. 

To apply the HV02 algorithm, it is required 
first to set the values of two parameters: the wall/field criterion $l$ and 
the minimum void-size threshold $s_{c}$. The value of the wall/field criterion 
was calculated as $l=\bar{d}_{3}+3\sigma_{3}/2$, where $\bar{d}_{3}$ is the 
mean distance to the third nearest neighbor halo and $\sigma_{3}$ is its 
standard deviation. If is found that $l=2.78,\ 2.70,\ 2.67$, and 
$2.75h^{-1}$Mpc, at $z=0,0.5,1$ and $2$, respectively.
To determine the size threshold, we tested the statistical significance
\citep{ela-pir97} that is defined
as $P(s_{c})=1-N_{\rm gap}(s_{c})/N_{\rm void}(s_{c})$ where
$N_{void}(s_{c})$ and $N_{gap}(s_{c})$ are the numbers of the voids found
in the real sample and in $10$ independent random samples, respectively.
This test basically allows us to discriminate the true voids from the Poisson
gaps. Fig.~\ref{fig:significance} plots $P(s_{c})$ as a function of the
size threshold, $s_{c}$ at $z=0,\ 0.5,\ 1$ and $2$ (solid, dashed, dot-dashed
and dashed line, respectively).  Noting that when $s_{c}=5.5$$h^{-1}$Mpc,
$P(s_{c})$ reaches $0.98$ at all redshifts, we set the void size threshold 
at $5.5$$h^{-1}$Mpc.

Table \ref{tab:void} lists the statistical properties of the voids
at each redshift. It is worth mentioning here that these properties
will depend on the void-finding algorithm. According to \citet{col-etal08}, 
the HV02 algorithm appears to identify as the void a much larger region 
(and thus identify many more galaxies as void-galaxies) than most of the other
algorithms. Thus, the mean effective comoving radius $\bar{R}_{v}$ and the 
number of void galaxies $N_{v}$ listed in Table \ref{tab:void} are expected 
to be larger than for the cases of most of the other algorithms. It also  
indicates that the number of void filaments and their sizes would be larger 
when the HV02 algorithm is used.  
\begin{table}
\centering \caption{The total number of the Millennium voids
$N_{v}$, the mean density contrast $\bar{\delta}_{v}$, and the mean
effective radius $\bar{R}_{v}$ at $z=0,0.5,1$ and $2$.}
\begin{tabular}{@{}cccc}
\hline
$z$& $N_{v}$ & $\bar{\delta}_{v}$ & $\bar{R}_{v}$\\
\hline
$0$   & $29186$ & $-0.90$ & $9.51$\\
$0.5$ & $27652$ & $-0.90$ & $9.42$\\
$1$   & $26615$ & $-0.89$ & $9.39$\\
$2$   & $27951$ & $-0.89$ & $9.42$\\
\hline
\end{tabular}
\label{tab:void}
\end{table}
\begin{figure*}
\begin{center}
\includegraphics[width=175mm]{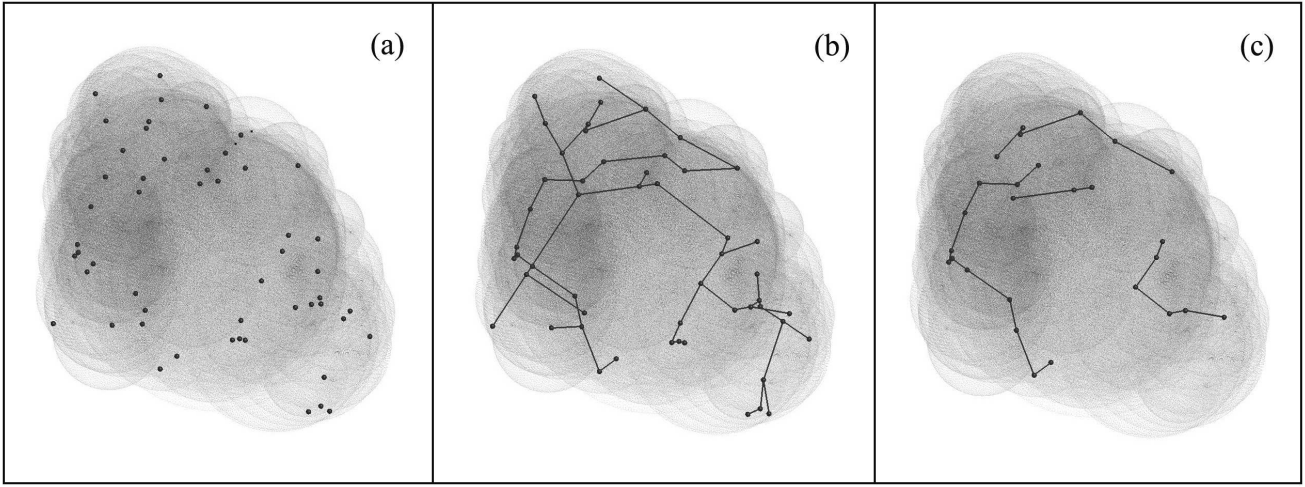}
\caption{(a) A two-dimensional projected image of a Millennium void
with dark matter halos (dots); (b) The minimal spanning tree (MST)
constructed from the void halos; (c): The void mini-filaments after the 
MST-reduction process.}
\label{fig:illustration}
\end{center}
\end{figure*}

\section{THE VOID FILAMENTS IN A $\Lambda$CDM UNIVERSE}
\label{filament}

We employ the filament-finding algorithm of \citet{bar-etal85} that
is based on the minimal spanning tree (MST) scheme to extract the
intrinsic linear patterns from the spatial distribution of the halos
in the selected large voids which contain more than $30$ halos. 
There are three key concepts for the MST scheme: node, edge and 
\textit{k}-branch. A node means the point (i.e., halo), an edge means a 
straight line that connects two nodes, and  a \textit{k}-branch means a 
path consisting of \textit{k}-edges with a leaf node only at one end. 
That is, one end of a \textit{k}-branch is free while the other end is 
an intersection of the edges. The MST of $N$ data points represents a 
unique network connected by $N-1$ edges of minimum total length without 
containing a closed loop.

To construct MSTs from the spatial distribution of the void halos,
we follow the simplest Prim process: First, an arbitrary node (halo) is
chosen as a starting point. Then the starting point is connected to its
nearest neighbor along a straight line (edge). The edge composed of two
nodes represents the first partial tree. The next nearest node is determined 
and added to the partial tree with an edge, via which the partial tree is 
extended.  After all nodes are connected to the tree according to the above 
prescription, the MST construction is completed.
The dominant linear patterns, i.e. filaments, are now extracted from
the MSTs by taking the following MST-reduction steps \citep{bar-etal85}:
\begin{enumerate}
\item \textit{Pruning}: An MST is pruned to the \textit{p} level by eliminating
all the \textit{k}-branches with \textit{k}$\leqslant$\textit{p}.
\item \textit{Separating}: All the edges longer than a given cut-off length
$l_{c}$ are removed.
\end{enumerate}
The \textit{Pruning} process allows us to find the main stems, removing 
the minor twigs that hardly contribute to the main structural patterns.
In other words, the superfluous small-scale noises from the MSTs are
minimized and the prominent filamentary patterns are highlighted
by the Pruning process. On the other hand, the Separating process breaks
apart the MSTs into distinct pieces, cutting off unphysical long linkages.
Both the pruning level $p$ and the cut-off length $l_{c}$ quantify the 
degree of the MST-reduction.

In most of the previous works where the large-scale filaments were the
targets to find by means of the MST technique, the values of $p$ and
$l_{c}$ were empirically determined as $p=9$ or $10$
\citep{bar-etal85, bha-lin88a, bha-lin88b} and $l_{c}=f\bar{l}$
\citep{bar-etal85,bha-lin88b,pli-etal92,pea-col95,krz-sas96,col-etal98}
where $\bar{l}$ is the mean edge-length of the unreduced
tree. And there were some authors who preferred $l_{c}=\bar{l}
+2\sigma_{l}$ or $l_{c}=\bar{l} +\sigma_{l}$
\citep{bha-lin88a,zuc-etal91}.
Unlike the previous works, however, our target is not the large-scale
filaments but the mini-filaments in void regions.  Accordingly, the relevant 
values of $p$ and $l_{c}$ are supposed to be different in our case. 
As a practical strategy to determine the values of $p$ and $l_{c}$ for the 
void filaments, we investigate how the final result (i.e., the size 
distribution of void filaments) changes with $p$ and $l_{c}$ and look for 
those values at which the final result stabilizes.

With the help of this strategy, we set $p=4$ and $l_{c}=\bar{l} 
+\sigma_{l}$ for the void filaments. Here, the range of $p$ is obtained 
empirically from the distribution of \textit{k} values of the unreduced 
trees. By setting $p=4$, we use only those branches which consist 
of at least $5$ edges as the main stems among the \textit{k}-branches. 
The detailed justification of $p=4$ is given in \S \ref{result}. 
As for $l_{c}=\bar{l} +\sigma_{l}$, we examine plenty of the 
Millennium voids and find that the Millennium void halos have maximum number 
of branches when $l_{c}=\bar{l} +\sigma_{l}$ 
\citep{gra-etal95,bas-etal07}.
\begin{figure*}
\begin{center}
\includegraphics[width=175mm]{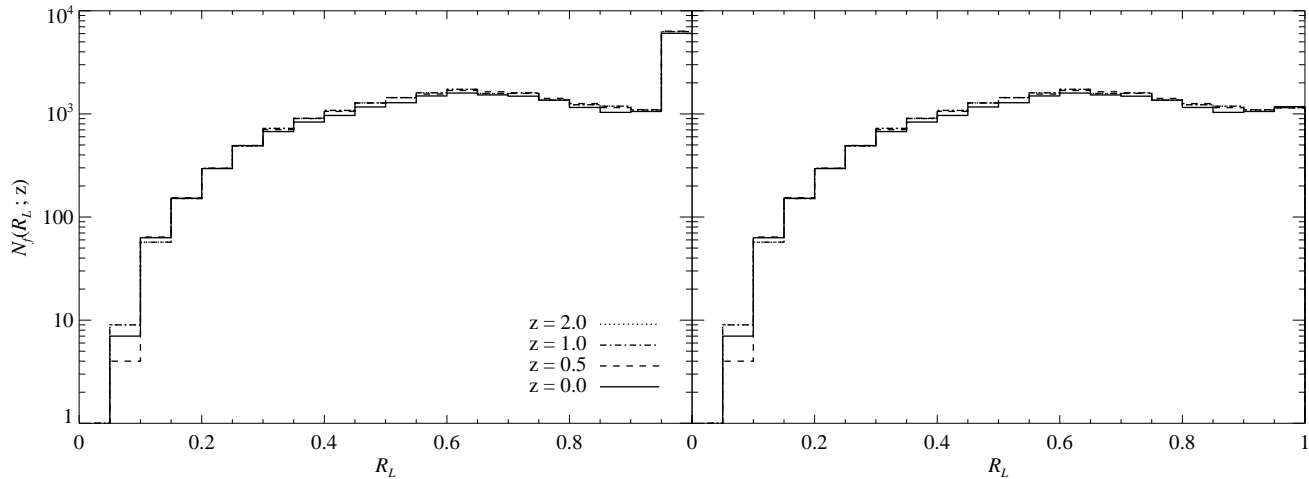}
\caption{Number distribution of void filaments as a function of the linearity, 
$R_{L}$ at $z=0,\ 0.5,\ 1,$ and $2$ as solid, dashed, dot-dashed, and dotted 
line, respectively, with single edges included (left) and excluded (right).} 
\label{fig:dis_Rl}
\end{center}
\end{figure*}
After completing the MST-reduction process, we end up with a final sample 
of the void filaments (reduced branches) with more than  one edges. 
Fig.~\ref{fig:illustration} depicts the spatial distribution of the 
mini-filaments in a Millennium void on a two dimensional projected plane.
Table~\ref{tab:filament} lists the statistical properties of the
void-filaments. Here, the linearity of a filament, $R_{L}$, is
defined by the ratio of the straight distance between the two end
nodes of the filament (end-to-end separation) to the sum of the lengths 
of all the edges constituting the filament \citep{bar-etal85}. Basically, 
it represents the degree of the straightness of a filament: 
The closer to unity the value of $R_{L}$ is, the more straight a given 
filament is.
\begin{table}
\centering \caption{The total number of voids with more than $30$
halos ($N_{v}$), the total number of void filaments ($N_{f}$), the range of 
the filament numbers per a void ($n_{\rm max}$), the mean number 
of filaments per a void ($\bar{n}_{f}$), the mean length of void filaments 
($\bar{R}_{f}$) in unit of $h^{-1}$Mpc, and the mean linearity of void 
filaments ($\bar{R}_{L}$) at $z=0,\ 0.5,\ 1,$ and $2$.}
\begin{tabular}{ccccccc}
\hline 
$z$ & $N_{v}$ & $N_{f}$ & $n_{f}$ & $\bar{n}_{f}$ 
& $\bar{R}_{f}$ & $\bar{R}_{L}$\\
 & & & & & $[h^{-1}$Mpc] & \\
\hline
$0$   & $5604$ & $22697$ & $[1,22]$ & $4$ & $15.06$ & $0.72$\\
$0.5$ & $5992$ & $24130$ & $[1,16]$ & $4$ & $14.79$ & $0.72$\\
$1$   & $6039$ & $24095$ & $[1,16]$ & $4$ & $14.69$ & $0.72$\\
$2$   & $5349$ & $20514$ & $[1,16]$ & $4$ & $15.11$ & $0.71$\\
\hline
\end{tabular}
\label{tab:filament}
\end{table}

Figure \ref{fig:dis_Rl} plots the number distribution of void filaments, 
$N_{f}(R_{L};z)$ vs. the linearity $R_{L}$ at $z=0,\ 0.5, \ 1$  and $2$ as 
solid, dashed, dot-dashed, and dotted histogram, respectively, with 
single-edge filaments included (left) and  excluded (right). As can be seen, 
when the single-edge filaments are included, the distribution has a peak 
around unity. In contrast, when the single-edge filaments are excluded, 
the peak occurs at $R_{L}=0.63$ and the mean linearity is reduced to 
$\bar{R}_{L}=0.64$.  This indicates that except for the single-edge filaments 
most of the void filaments are not straight but curved. It can be readily 
understood since the void filaments are usually located in the boundary of 
voids.

\begin{figure}
\begin{center}
\includegraphics[width=84mm]{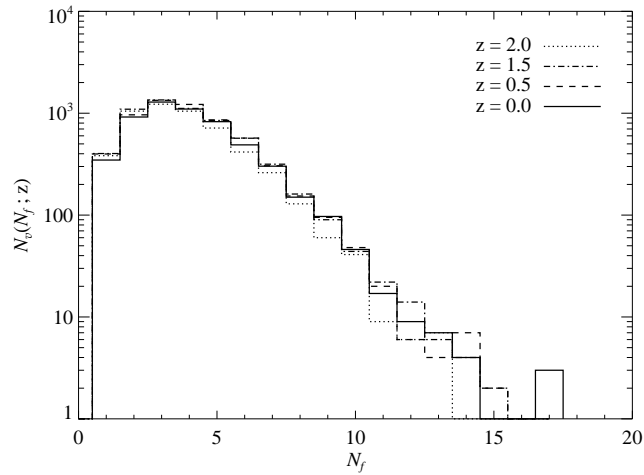}
\caption{Number distribution of voids as a function of the number of their 
member filaments at $z=0,\ 0.5,\ 1,$ and $2$ as solid, dashed, dot-dashed, 
and dotted line, respectively.} \label{fig:dis_nf}
\end{center}
\end{figure}

Figure~\ref{fig:dis_nf} plots the number distribution of voids,
$N_{v}(N_{f};z)$, as a function of the number of member filaments,
$N_{f}$, at $z=0,\ 0.5, \ 1$ and $2$ as solid, dashed, dot-dashed,
and dotted histogram, respectively. Most of the voids have less than 
$10$ member filaments.

\section{THE SIZE DISTRIBUTION OF VOID MINI-FILAMENTS} \label{result}

\begin{figure*}
\begin{center}
\includegraphics[width=175mm]{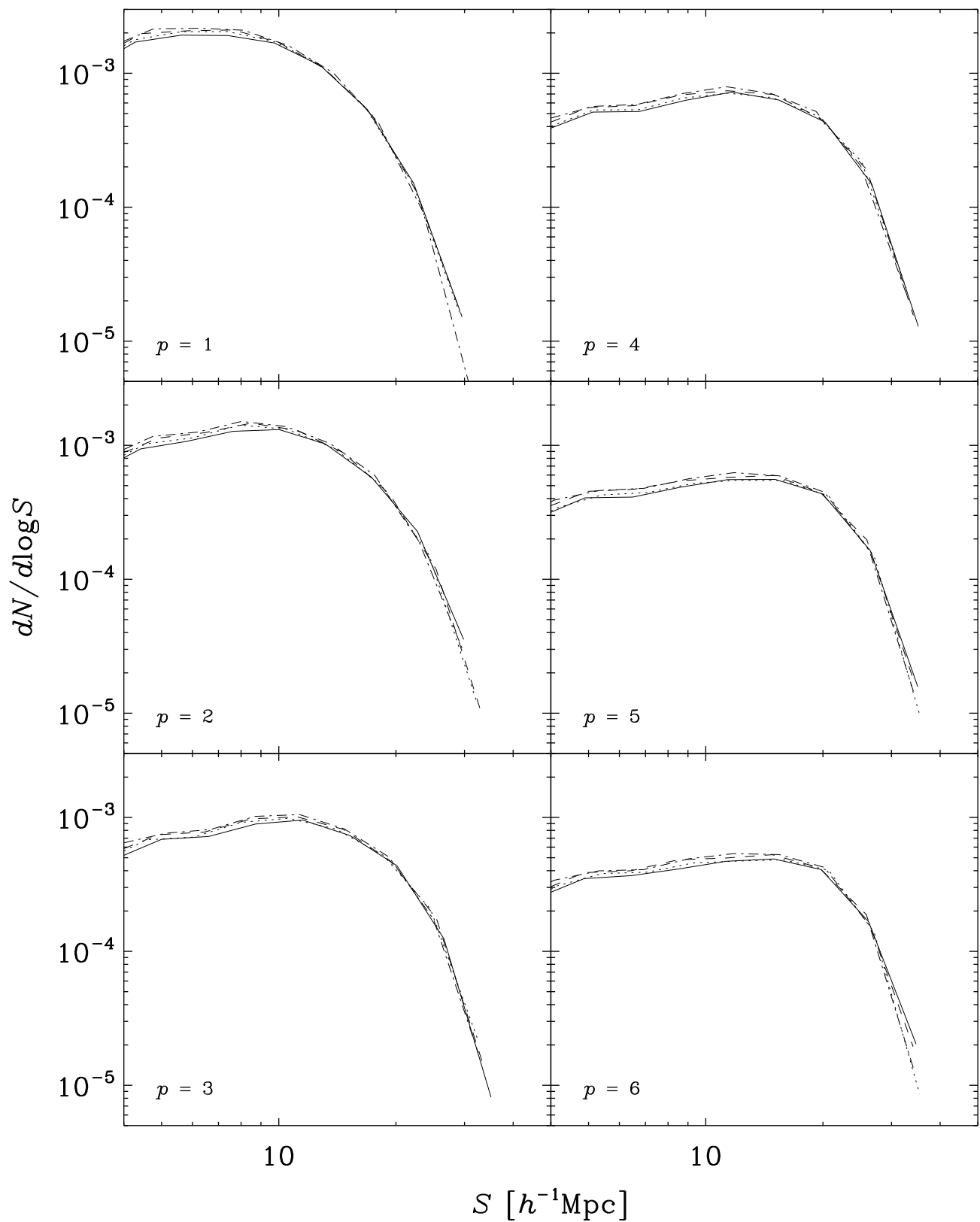}
\caption{Number density of void-filaments per unit volume in a logarithmic 
size bin for the six different cases of the pruning level, $p$ at $z=0,\ 
0.5,\ 1$ and $2$ (solid, dashed, dot-dashed, and dotted line, 
respectively).}
\label{fig:size_dis}
\end{center}
\end{figure*}

We first adopt the definition of \citet{col07} for the filament size, $S$, 
and measure $S$ of a given filament as the length of the diagonal line of 
a cuboid that exactly fits the spatial extent of the filament nodes.
Let $\left(x_{\min}, x_{\max}\right)$, $\left(y_{\min},y_{\max}\right)$, 
and $\left(z_{\min}, z_{\max}\right)$ represent the full ranges of the x, 
y and z positions of the filament nodes, respectively. 
Then, $S$ is determined as
\begin{equation}
\label{eqn:size}
S = \sqrt{\left({x_{\max }-x_{\min }}\right)^{2}+
\left({y_{\max }-y_{\min }}\right)^{2}  + 
\left({z_{\max }-z_{\min}}\right)^{2}}.
\end{equation}
By equation (\ref{eqn:size}), we measure the size of every void filament 
and determine the number density of void filaments as a function of $\log S$.
Fig.~\ref{fig:size_dis} plots the size distribution of void filaments,
$dN/d\log S$, for the six different cases of the pruning level $p$ at
$z=0,\ 0.5,\ 1$, and $2$ (solid, dashed, dot-dashed, and dotted line, 
respectively).
When $p$ increases from $1$ to $3$, there are noticeable changes in the shape
of $dN/d\log S$. Whereas, when $p$ increases from $4$ to $6$, there is only
little change in the shape of $dN/d\log S$. It indicates that the onset of
the stabilization of $dN/d\log S$ occurs at $p=4$.  No branches are further 
pruned and thus the main structures of the MSTs are retained after $p=4$.
This trend justifies our choice of the pruning level $p=4$ in \S 3.
As can be seen in Fig.~\ref{fig:size_dis}, the distribution $dN/d\log S$ 
appears insensitive to $z$. The four lines in each panel behave very 
similarly, which implies that it may be possible to express $dN/d\log S$ 
in a universal form.  Given the fact that the sizes of the void filaments 
depend on the sizes of their host voids, we use the rescaled filament size, 
$\nu\equiv S/R_{v}$ where $R_{v}$ is the effective comoving radius of a host 
void that is calculated by means of the Monte Carlo integration method 
as described in HV02 \citep[see also][]{par-lee07b}.

\begin{figure}
\begin{center}
\includegraphics[width=84mm]{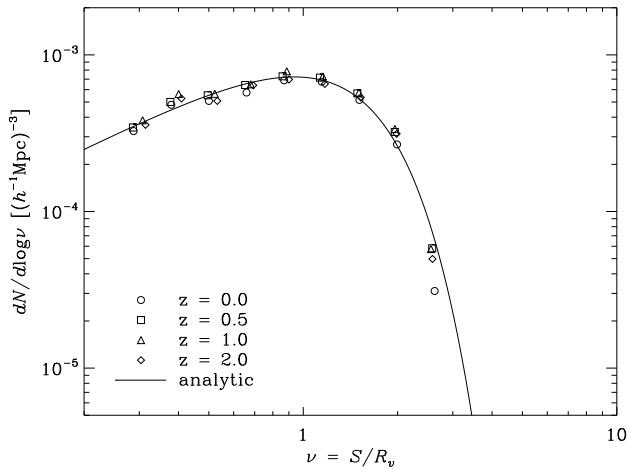}
\caption{Number density of void filaments as a function of the rescaled 
size at $z=0,\ 0.5,\ 1$ and $2$ (open circles, squares, triangles, and 
diamonds, respectively). The solid line represents a fitting function 
(eq.[\ref{eqn:fit1}]) with the best-fit values of $A$ and $B$.
The filament size is measured as the spatial extent of the filament nodes.}
\label{fig:SE}
\end{center}
\end{figure}

It is found that the rescaled size distribution of void filaments 
$dN/d\log\nu$ at all redshifts can be well fitted to the following analytic 
formula:
\begin{equation}
\label{eqn:fit1}
f(\nu)  = A\nu \exp \left( {-B\nu^2 } \right).
\end{equation}
where $A$ and $B$ are the fitting parameters whose best-fit values are
found through the $\chi^{2}$-minimization. Fig.~\ref{fig:SE} plots the 
rescaled size distributions of void filaments at four redshifts and 
compares them with the analytic fitting formula with the best-fit parameters.
As can be seen, the numerical results at four redshifts are indeed in good 
agreement with the fitting model.

\begin{figure}
\begin{center}
\includegraphics[width=84mm]{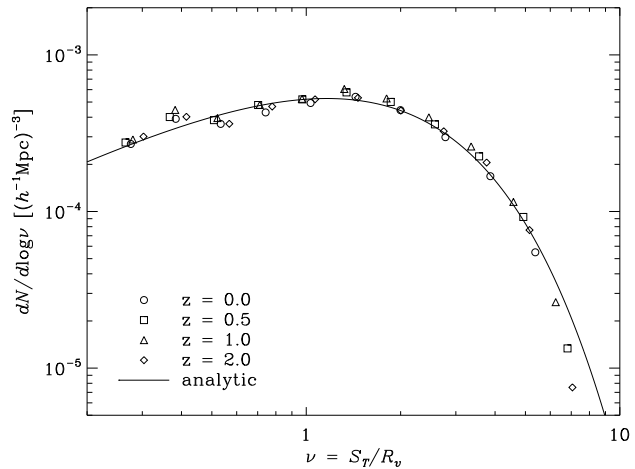}
\caption{Same as Fig.\ref{fig:SE}, but the sizes of void filaments
are measured as the sum of the lengths of all edges that constitute 
the filaments, $S_T$.}
\label{fig:TL}
\end{center}
\end{figure}

Yet, a careful reader may well suspect that the universality and the 
functional form of $dN/d\log\nu$ may depend on the definition of the filament 
size. To test how the fitting model and the behavior of $dN/d\log\nu$ changes 
with the definition of the filament size, we redefine the size of a filament 
as its total length, $S_{T}$, (the sum of the lengths of all edges that 
constitute the filament) and repeat the whole process. It is found that 
$dN/d\log\nu$ is still almost universal but the functional form of the 
fitting model is different.
\begin{equation}
\label{eqn:fit2}
f(\nu) = A\nu\exp\left(-B\nu\right).
\end{equation}
Fig.~\ref{fig:TL} plots the same as in Fig.~\ref{fig:SE} but for the case 
that the filament size is measured as the total length, $S_{T}$. As can be 
seen, in this case $dN/d\log\nu$ decreases less rapidly. It is because the 
total length of a given filament is usually longer than its spatial extent.  
The more convoluted a filament is, the larger the difference between $S_{T}$ 
and $S$ is. 

To make a more quantitative test of the universality of $dN/d\log\nu$, we 
conduct a $\chi^{2}$-statistics.  Assuming the numerical result of 
$dN/d\log\nu$ at each redshift as one independent realization, we calculate 
the one standard deviation of $dN/d\log\nu$ between the four realizations, 
$\sigma_{j}$.  One scatter error $\sigma$ in the measurement of universal 
value is calculated as $\sigma\equiv \sqrt{\sigma^{2}_{j}+\sigma^{2}_{p}}$ 
where $\sigma^{2}_{p}$ denotes the Poisson errors. With this error at each 
bin, we evaluate the reduced $\chi^{2}$, 
$\chi^{2}_{r}\equiv \chi^{2}/n_{\rm f}$ ($n_{\rm f}$: the degree of freedogm) 
averaged over all size bins. The values of $\chi^{2}_{r}$ are listed in 
Table \ref{tab:fit} where the best-fit values of $A$ and $B$ can be also 
found for the cases of two different definitions of $S$.  
As can be seen, the values of $\chi^{2}$ are reasonably close to unity, which 
proves that the rescaled size distribution of void filaments is almost 
universal. But, it is worth noting that when the filament size is defined 
as the spatial extent, the distribution is more insensitive to redshift.
\begin{table}
\centering \caption{The results of the reduced $\chi^{2}$ test for
the size distribution of void filaments for the two different cases of 
the filament-size definition.}
\begin{tabular}{@{}cccc}
\hline
size & $A$ & $B$ & $\chi^2$\\
\hline
spatial extent  & $0.00127$ & $0.569$ & $1.71$\\
total length & $0.00123$ & $0.859$ & $2.05$\\
\hline
\end{tabular}
\label{tab:fit}
\end{table}

\section{DISCUSSION}  \label{end}

It is generally thought that the large scale structure of the universe
provides a window on the initial condition of the early universe.
In previous cosmological studies,  it was the halo mass function that has
been most highlighted as a statistical tool for probing cosmology with the
large scale structure
\citep{pre-sch74,she-tor99,she-etal01,jen-etal01,ree-etal03}.
The most advantageous aspect of the halo mass function is that it
can be written in a simple universal form, independent of redshift
\citep{she-tor99}. Yet, the halo mass function has a
generic weakness as a cosmological probe: its dependence on the
cosmological parameters comes indirectly from the dependence of the
halo mass on the rms fluctuation, $\sigma(R)$, of the linear density field
smoothed on a scale radius of $R$.

It is desirable to develop another statistical tool which can
overcome the weakness of the halo mass function. Recently, several
authors have noted the possibility of probing cosmology with cosmic
voids \citep[e.g.,][]{she-van04,par-lee07a,pla-etal08}. Yet, no new
statistical tool that can really compete with the halo mass function
has been suggested so far. For instance, in our previous work
\citep{par-lee07a}, we have proposed that the cosmic voids provide
another window on cosmology. Given that the void ellipticities are
induced by the primordial tidal effect which depends on the
background cosmology, we have shown analytically that the void
ellipticity distribution can be used to constrain the cosmological
parameters.

However, it has turned out that the void ellipticity distribution
suffers from the same weakness that the halo mass function has. Its
dependence on cosmology is secondary, resulting from the dependence
of the void scale on the linear rms density fluctuation, $\sigma(R)$. 
Furthermore, the void ellipticity distribution cannot be written in a 
universal form unlike the halo mass function. Therefore, true as it is 
that the void ellipticity distribution provides another way to determine 
the cosmological parameters with the large scale structure, it is not 
competitively compared with the halo mass function.

In this work, we have for the first time quantified the effect of the tidal
field on cosmic voids by the sizes of the void filaments. It is shown that
the size distribution of void filaments is almost independent of redshift,
having a simple universal form. A crucial implication of our result is that
the number density of long filaments in voids should depend very sensitively
on the background cosmology since the sizes of the void filaments reflect
the primordial tidal field.

Our result that the size distribution of void filaments is almost independent
of redshift in spite of the well known fact that the sizes of voids grow very
strongly with redshift can be understood as follows.
It is true that the comoving sizes of voids keep growing as the voids expand
faster than the rest of universe due to their extreme low density. However,
the comoving sizes of void filaments should not necessarily keep growing with
redshift.

But for the effect of the tidal field, the spatial distribution of
void halos would be isotropic due to the gravitational rarefaction
effect caused by the fast expansion of voids. In other words, while
the gravitational rarefaction effect tends to make the spatial
distribution of void halos isotropic, the tidal effect on voids
tends to increase the anisotropy in the distribution of void halos,
resulting in growth of void filaments. Therefore, as far as the
tidal effect exists and counteracts the expansion effect, the sizes
of void filaments will also grow as the sizes of their host voids
grow with redshift.

However, when the void size grows large enough to finally reach its
threshold at which the expansion effect overcomes the tidal effect,
the degree of anisotropy in the spatial distribution of void halos
will diminish and accordingly the sizes of void filaments will stop
growing. Since this critical threshold of void size is determined by
the single condition that the expansion effect overcomes the tidal
effect, its value should be redshift-independent.  At high-redshift
when the sizes of voids were small, the strong dominant tidal effect
increases the anisotropy in the spatial distribution of void halos,
increasing the sizes of void filaments. At low-redshift when the
sizes of voids are large, the strong expansion effect prevents the
sizes of void filaments from keep growing. Hence, the universal size
distribution of void filaments reflects the fact that the sizes of
void filaments are regulated by the counter-balance between the
expansion effect and the tidal effect.

The fitting formula, eqs. (\ref{eqn:fit1})-(\ref{eqn:fit2}), indicate 
that the number density of void filaments, $dN/d\log\nu$, behaves 
like a power-law with power index of $n=1$ in the short filament section
while it decreases exponentially in the long filament section. 
A crucial implication of our result is that the exponential decrease
of the abundance of long-size filaments in voids should be very
sensitive to the key cosmological parameters. Especially it is
expected to depend sensitively on the amplitude of the linear power
spectrum, $\sigma_{8}$. In our previous work, we have already shown
that the void ellipticity caused by the tidal effect increases with
the value of $\sigma_{8}$ \citep{par-lee07a}. Accordingly, we expect
that the number density of long filaments in voids would increase as
the value of $\sigma_{8}$ increases. Unlike the void ellipticity
distribution whose dependence on $\sigma_{8}$ is weak and indirect
through its dependence on the rms density fluctuation, equation
(\ref{eqn:fit1}) suggests that the size distribution of void
filaments should depend strongly on the power spectrum amplitude.

To use the size distribution of void filaments as a cosmological probe,
however, there should be some additional tasks to be done in the future.
First, the mass-to-light bias and the redshift distortion effects have
to be account for. What one can observe is not halos in real space but
galaxies in redshift space. The size distribution of void filaments measured
from the galaxy catalogs in redshift space could differ from the current
result. Thus, it will be very necessary to investigate how the bias and the
redshift distortion effect change the size distribution of void filaments.
It is expected that the sizes of void filaments may increase in redshift 
space since the redshift space distortion effect tend to increase the 
anisotropy in the spatial distribution of void galaxies. Meanwhile, the 
matter-to-light bias might decrease the sizes of void filaments, compensating 
for the redshift distortion effect, since the massive halos are found to 
be distributed less anisotropically (Park \& Lee 2009 in preparation). 
Henceforth, the overall size distribution of void galaxy filaments in 
redshift space might be similar to that in real space, due to the competition 
between the two effects (in private communication with van de Weygaert).

Second, a general analytic formula for the size distribution of void filaments
in an arbitrary cosmology is desirable to derive from physical principles.
If an analytic model is found, it would allow us to understand the true
physical meaning of eqs. (\ref{eqn:fit1})-(\ref{eqn:fit2}). 
Third, it should be worth examining whether the results depend on the choice 
of the filament-finding and the void-finding algorithms. As mentioned in 
\S 1, it has been found that different void-finders provide different void 
properties and the HV02 algorithm that we used here has a tendency to 
find larger voids than most of the other void-finders \citep{col-etal08}.
Furthermore, our results are not general ones but valid only for the specific 
cosmology of the Millennium Run simulation. Therefore, it is required to test 
the universality of the size distribution of void filaments with other 
void-finders and with other cosmologies as well. We plan to address these 
issues and report the results elsewhere in the near future.

\section*{Acknowledgments}

The authors would like to thank a referee for a constructive report.
The Millennium Run simulation used in this paper was carried out by the
Virgo Supercomputing Consortium at the Computing Center of the Max-Planck
Society in Garching. The Millennium Simulation data are available at 
http://www.mpa-garching.mpg.de/millennium.
This work is financially supported by the Korea Science and Engineering 
Foundation (KOSEF) grant funded by the Korean Government
(MOST, NO. R01-2007-000-10246-0).


\end{document}